\documentclass[twocolumn,showpacs,preprintnumbers,amsmath,amssymb]{revtex4}

\usepackage{graphicx}
\usepackage[dvips]{color}

\begin{document}


\title{Radio-frequency transitions on weakly-bound ultracold molecules}

\author{Cheng Chin$^1$ and Paul S. Julienne$^2$}

\affiliation{$^1$Institut f\"{u}r Experimentalphysik,
Universit\"{a}t Innsbruck, Technikerstr. 25, 6020 Innsbruck,
Austria\\
$^2$Atomic Physics Division, National Institute of Standards and
Technology\\100 Bureau Drive Stop 8423, Gaithersburg, Maryland
20899}

\date{\today}

\begin{abstract}
We show that radio-frequency spectroscopy on weakly-bound
molecules is a powerful and sensitive tool to probe molecular
energy structure as well as atomic scattering properties. An
analytic expression of the rf excitation lineshape is derived,
which in general contains a bound-free component and a bound-bound
component. In particular, we show that the bound-free process
strongly depends on the sign of the scattering length in the
outgoing channel and acquires a Fano-type profile near a Feshbach
resonance. The derived lineshapes provide an excellent fit to both
the numerical calculation and the experimental measurements.
\end{abstract}



\maketitle\narrowtext\section{Introduction}

Radio-frequency (RF) spectroscopy is widely applied to many
experiments on atoms or molecules, for which an exquisite energy
resolution can be achieved. This is due to the magnetic coupling
nature of the RF transition and the associated long coherence time
of the systems. In recent experiments on ultracold weakly-bound
molecules, rf spectroscopy also allows a precise determination of
the molecular binding energy \cite{debbieRF} and the pairing gap
in a degenerate Fermi gas \cite{gap}. In the latter case, the high
energy resolution reveals the fermionic nature of the pairing in
the Bardeen-Cooper-Schrieffer (BCS) regime.

In this article we investigate theoretically the radio-frequency
excitation spectrum of ultracold molecules. Previous experimental
work in this regime shows that molecules are dissociated upon
receiving the RF photons \cite{debbieRF, gap}. The associated
bound-free excitation lineshape is characteristically broader and
highly asymmetric compared to that for atoms. Here, we provide an
intuitive picture to model and derive a simple analytic formula
for the excitation rates and the lineshapes. We show that much
information regarding cold collision properties can be extracted
from the lineshape. In particular, the spectrum is very sensitive
to tuning a Feshbach resonance in the final state scattering
continuum, where the rf spectra will evolve into two components: a
bound-bound transition to the newly formed molecular state and a
remaining weaker bound-free transition.

Our analytic results provide excellent fits to both the numerical
calculations and the experimental measurements from the Innsbruck
group \cite{gap, boundbound}. We argue that radio-frequency
spectroscopy on cold molecules has several advantages in probing
ultracold collision properties over the conventional collision
measurements. These include a high energy resolution, the
flexibility to probe different channels and its insensitivity to
the sample density and temperature.

In this paper, we first introduce our model and derive the
lineshape (Sec. II). We then compare our results to the numerical
calculation (Sec. III) and discuss the spectral feature (Sec. IV).
Finally, we compare our results to the experimental data (Sec.
IV).

\section{Model}
We consider a weakly-bound molecule in the state $|m\rangle$. The
bound state consists of two atoms with binding energy
$E_b=\hbar^2/2\mu(a-r_0)^2>0$ relative to the dissociation
continuum $A$~\cite{flambaum}. Here $\hbar$ is Planck's constant, $\mu$
is the reduced mass of the two atoms, $a$ is the scattering length in the
scattering channel $A$ and $r_0$ is the interaction range
of the van der Waal potential, which varies with interatomic separation $r$
as $-C_6/r^6$:
\begin{equation}
  r_0 =  2^{-3/2} \frac{\Gamma\left(\frac{3}{4}\right)}{\Gamma\left(\frac{5}{4}\right)}
  \left( \frac{2 \mu C_6}{\hbar^2} \right)^{1/4} \,. \label{r0}
\end{equation}
For weakly-bound molecules, we assume $a\gg r_0$. The molecule is
initially at rest and a radio-frequency photon with energy
$E_{RF}$ couples the molecule to a different channel $A'$,
characterized by the scattering length $a'$. For $a'<0$, the final
state is a continuum and the excited molecule dissociates; for
$a'>0$, a stable bound state $|m'\rangle$ is also available in the
final state channel and the molecule can either dissociate or be
driven to the bound state $|m'\rangle$. In the following, we
assume that both scattering channels are in the threshold regime
($a$, $|a' |\gg r_0$). Atoms, bound or unbound, then have the same
internal wavefunction. We also assume the relative energy of the
continuum threshold to be $E_{A'}-E_A=E_0>0$.

\subsection{Bound-free transition}

First, we consider the bound-free transition. Energy conservation
gives
\begin{eqnarray}
E_{RF}=E_0+E_b+K,  \label{erf}
\end{eqnarray}
where $K=\hbar^2 k^2/2\mu>0$ is the kinetic energy of the outgoing
wave and $k$ is the associated wavenumber. For $E_{RF}<E_0+E_b$,
the transition is forbidden. For $E_{RF}\geq E_0+E_b$, the
bound-free transition rate from the initial state $|m\rangle$ to
the final state $|K\rangle$ is given by the Fermi's golden rule:
\begin{equation}
\Gamma_f(K)=\frac{2\pi}{\hbar}  \left | \left \langle K
\left | \frac{\hbar\hat{\Omega}}{2}
\right  | m \right \rangle \right |^2
 =\frac{h \Omega^2}{2} F_f(K) , \label{gfk}
\end{equation}
where
\begin{equation}
 F_f(K)= \left | \int \psi^*_{K}(r) \phi_m(r) dr \right |^2
 \label{ffk}
\end{equation}
is the bound-free Franck-Condon factor per unit energy,
$\hbar\hat{\Omega} /2$ is the RF interaction for the rf Rabi
frequency $\Omega$, $\phi_m(r)$ is the bound molecular
wavefunction in channel $A$, and $\psi_{K}(r)$ is the
energy-normalized s-wave scattering wavefunction in channel $A'$
\cite{energyeigen}.

We can evaluate the bound-free Franck-Condon factor $F_f$ based on
the asymptotic behavior of the wavefunctions.
\begin{eqnarray}
\psi_K(r)&=&\sqrt{\frac{2\mu}{\pi\hbar^2 k}} \sin{(kr+\delta')}
\label{psikr} \\
\phi_m(r)&=&\sqrt{\frac2a} e^{-r/a}, \label{phimr}
\end{eqnarray}
where $\delta'$ is the scattering phase shift in channel A'.
Combining Eqs.~ (\ref{ffk}), (\ref{psikr}) and (\ref{phimr}), we
have
\begin{eqnarray}
F_f(K)=\frac{4\mu a}{\pi \hbar^2 k} (1+k^2a^2)^{-2}
(\sin\delta+ka\cos\delta')^2  \,.\label{ffk2}
\end{eqnarray}

Since $k \ll 1/r_0$, we can use the low energy expansion of
the scattering phase shift \cite{mott},
\begin{eqnarray}
k\cot{\delta'}=-\frac{1}{a'}+\frac{r'_e}{2}k^2+O(k^4) \,,
\label{kcotd}
\end{eqnarray}
where the effective range $r'_e$ depends on the
scattering length $a'$ and the interaction range $r_0$. The explicit
form of $r'_e$ for a van der Waals potential is given in Ref.
\cite{gao} in the limit $|a'| \gg r_0$:
\begin{equation}
   r'_e = \frac{\Gamma(\frac{1}{4})^4}{6\pi^2} r_0 \,.
   \label{effrange}
\end{equation}

By taking the leading term in the expansion in Eq.~(\ref{kcotd})
and expressing $k^2 a^2=K/E_b$ and $k^2 {a'}^2=K/E_b'$, we derive
a simple and very useful form of the lineshape,
\begin{eqnarray}
F_f(K)=\frac{2}{\pi}\left (1-\frac{a'}{a}\right)^2
\sqrt{\frac{K}{E_b^3}}\left
(1+\frac{K}{E_b}\right)^{-2}\left(1+\frac{K}{E_b'}\right)^{-1}.
\label{ffkfinal}
\end{eqnarray}

In the limit $k\rightarrow 0$, where $\delta'=-ka'$, we have
$F_f\sim K^{1/2}$, which is the Wigner threshold regime. The
transition rate peaks between $K\sim 0$ for $|a'|\gg a$ and
$K=E_b/3$ for $|a'|\ll a$ and decreases to higher $K$. For very
large $K\geq \hbar^2/mr_0^2$, Eq.~(\ref{kcotd}) is no longer
valid. A few extreme situations including $a=a'$, where $E_f$
vanishes and $a'=\pm\infty$, where the Wigner threshold law fails,
will be discussed in later sections.

The integrated rf line strength $S_f$ is given by
\begin{eqnarray}
S_f&=&\int_0^{\infty}F_f(K)dK \\
   &=&\left\{\begin{array}{lr}
   1  & \mbox{for }a'\leq0 \\
   (\frac{a-a'}{a+a'})^2  & \mbox{for }a'>0
   \end{array}
   \right.
\label{bfStrength}
\end{eqnarray}
The unit Franck-Condon factor for $a'<0$ means that within the
approximations we made, the molecular wavefunction  $\phi_m(r)$ is
fully expanded by the scattering states $|K\rangle$. For $a'>0$,
we have $S_f<1$, which implies there are additional outgoing
channels. This is because the additional bound-bound transition
process is opened for $a'>0$.

\subsection{Bound-bound transition}

For positive scattering length in the outgoing channel $a'\gg
r_0>0$, a weakly-bound molecular state $|m\rangle$ exists.
Bound-bound transitions are allowed. From energy conservation, we
have
\begin{eqnarray}
E_{RF}=E_{0}+E_b-E_{b}',  \label{erf2}
\end{eqnarray}
Effectively, $K=-E_b'$. The bound-bound Franck-Condon factor
$F_b(K)$ can be calculated similar to Eqs.~(\ref{gfk}) and
(\ref{ffk}).
\begin{eqnarray}
F_b(K)&=&|\int\phi^*_{m'}(r)\phi_m(r)dr|^2\delta(K+E_b')\\
&=&\frac{4aa'}{(a+a')^2}\delta(K+E_b'), \label{fbk}
\end{eqnarray}
where we have used the molecular wavefunction in Eq.~(\ref{phimr})
for both $|m\rangle$ and $|m'\rangle$ states and have introduced
the $\delta$-function to provide energy normalization analogous to
that of $F_f(K)$. The bound-bound line
strength is then $S_b=4aa'/(a+a')^2$.

Note that sum of the bound-bound and bound-free integrated line strengths is
identically one for $a'>0$, as expected from the wavefunction
projection theorem:
\begin{eqnarray}
S_f+S_b=1 \,. \label{sum}
\end{eqnarray}

\section{Comparison with numerical calculation}



\begin{figure}
\includegraphics[width=2.75in]{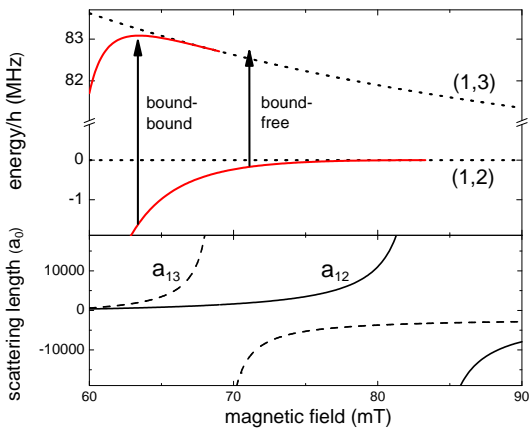}
\caption{Energy structure and the scattering lengths of $^6$Li$_2$
in the $(1,3)$ and $(1,2)$ channels (dotted lines). All energies
are referenced to the $(1,2)$ scattering threshold. In these two
channels, Feshbach couplings induce the formation of molecules
(solid lines) below 69.1mT and 83.5mT, respectively. Arrows show
the bound-bound and bound-free transitions based on molecules in
the $(1,2)$ channel. The lower figure shows the scattering lengths
in the two channels.} \label{fig1}
\end{figure}

We check the validity of the above analytic formulas by comparing them
to a numerical calculation. We choose fermionic $^6$Li atoms as
our model system, for which the interaction parameters are
precisely known by fitting various cold atom and molecule
measurements to a multichannel quantum scattering calculation
\cite{boundbound}. The energy structure and the scattering
lengths in the two relevant channels $A=(1,2)$ and $A'=(1,3)$ are
shown in Fig.~\ref{fig1}. Here, $(1,2)$ refers to the state with
one atom in state $|1\rangle$ and one in $|2\rangle$, and
$|N\rangle$ refers to the lowest $N$th internal state in the
$^6$Li atom ground state manifold.

\begin{figure}
\includegraphics[width=2.75in]{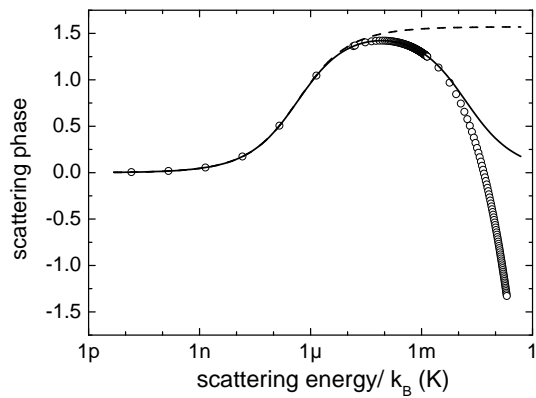}
\caption{Comparison of the scattering phase shifts from numerical
calculation and from the effective range expansion. The numerical
calculation (open circles) is shown together with the predictions
from Eq.~(\ref{kcotd}) with (solid line) and without (dashed line)
the effective range correction.
The scattering parameters are based on $^6$Li in the $(1,3)$
channel near 72.0mT with $a_{13}=-7866a_0$.} \label{fig2}
\end{figure}

To calculate the scattering phase shifts, we construct a reduced
single-channel Hamiltonian to describe the $(1,3)$ continuum. The
phase shifts from this model are nearly indistinguishable from
those from the full multichannel calculations in this range of
$B$-fields.  The van der Waals interaction is set to
$C_6=1393.39(16)$a.u. \cite{lic6}, yielding an interaction range
of $r_0=29.884(3)a_0$. The scattering lengths are set to match the
values from the multi-channel calculation. The resulting
scattering phase shifts from the calculation are compared with the
effective range expansion given in Eq.~(\ref{kcotd}) (see
Fig.~\ref{fig2}). Here, the scattering parameters are based on the
$(1,3)$ scattering states at 72.0mT with a scattering length of
$a'=-7866a_0$. The result shows the expected behavior that
$k\cot\delta'=-1/a'$ provides an excellent fit at low scattering
energy $E<k_B\times20\mu$ K and the effective range correction
works up to $E<k_B\times10$mK. Here $k_B$ is Boltzman's constant.

%
%





\begin{figure}
\includegraphics[width=2.75in]{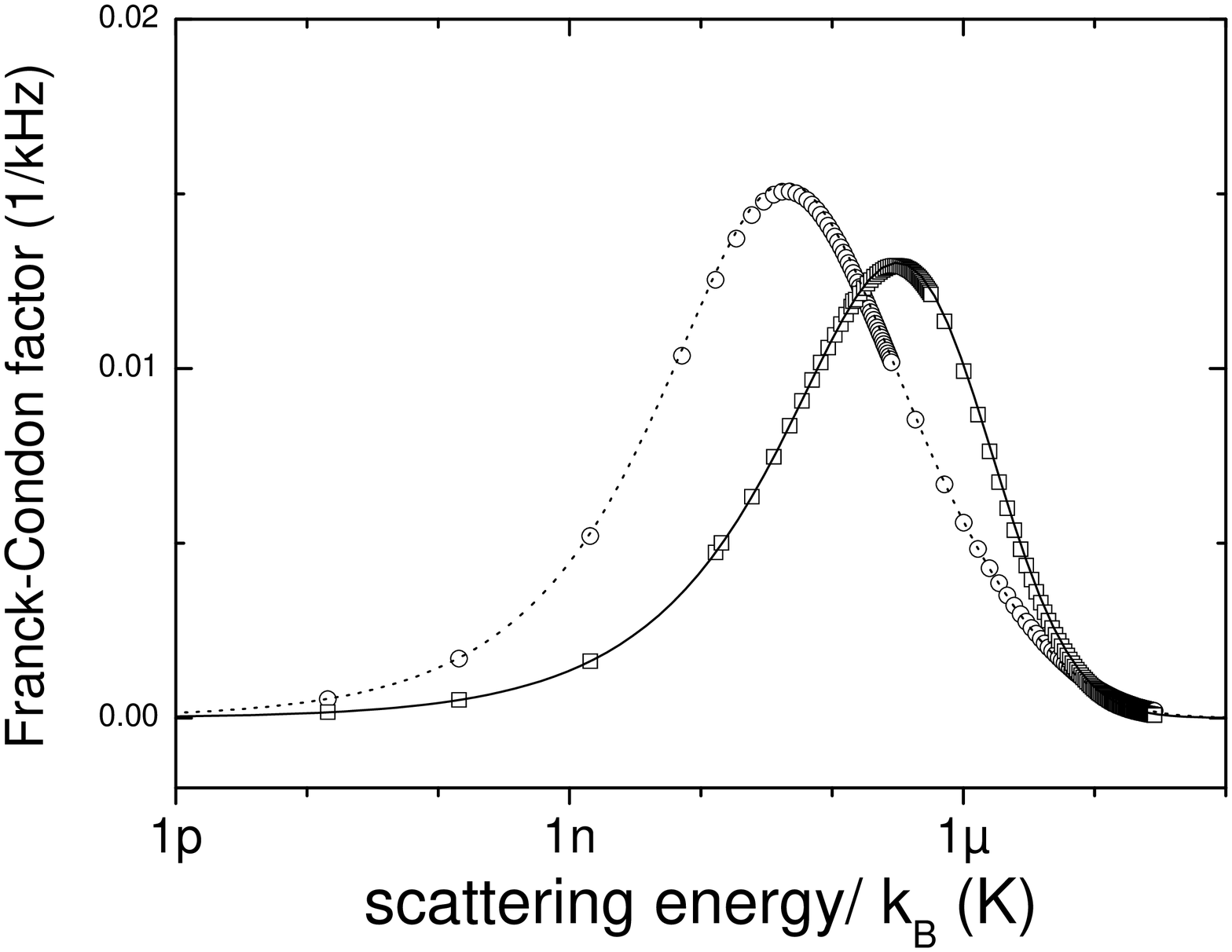}
\caption{Comparison of the Franck-Condon factors from numerical
calculation and from theory. Numerical calculation at 72.0mT (open
squares) and 68.0mT (open circles) are plotted together with the
formula Eq.~(\ref{ffkfinal}) (solid line at 72.0mT and dotted line
at 68.0mT). The parameters are $E_b=k_B\times 16.2\mu$K and
$a'=25173a_0$ at 68.0mT; $E_b=k_b\times 6.2\mu$K and $a'=-7866a_0$
at 72.0mT.} \label{fig3}
\end{figure}

We also compare the Franck-Condon factors obtained from the
numerical calculation and from Eq.~(\ref{ffkfinal}). The numerical
calculation is based on an initial bound state in the $(1,2)$
channel and the final state $(1,3)$ continuum, described by the
reduced Hamiltonian. We show the calculations at two magnetic
fields of 72.0mT and 68.0mT, where the scattering parameters are
set according to the multichannel calculations. The results show
that Eq.~(\ref{ffkfinal}) provides an excellent fit to the full
bound-free spectra, see Fig.~\ref{fig3}.

\section{RF lineshape near a Feshbach resonance}

The appearance of the bound-bound transition for only $a'>0$ seems
to suggest a distinct behavior in rf excitation near the Feshbach
resonance where the scattering length changes sign. In this section,
we show that  the evolution of the rf spectrum is actually continuous
when a $B$-field is tuned through the resonance position.
We use Eq.~(\ref{ffkfinal}) andEq.~(\ref{fbk}) to show the bound-free
and bound-bound spectra in the vicinity of a Feshbach resonance.

\begin{figure}
\includegraphics[width=2.75in]{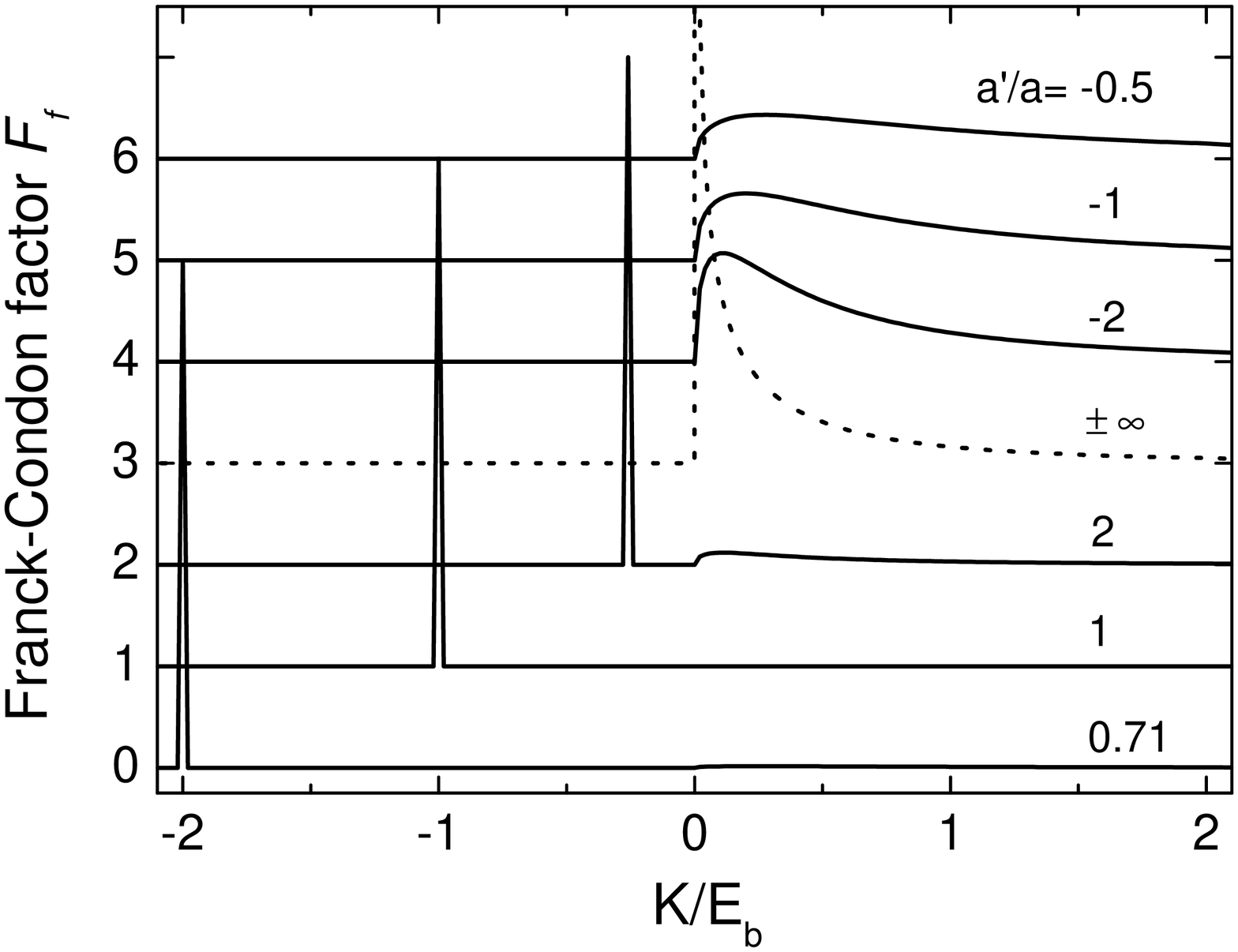}
\caption{Transition from a bound-free transition to a bound-bound
transition with $E_b=1$. For the bound-free transition, the
Franck-Condon factor is plotted as a function of $K/E_b$. Curves
of different $a'/a$ are offset and shown in the order of that near
a Feshbach resonance in the $A'$ channel. (dotted line for the
resonance condition) For $a'>0$, the locations of the bound-bound
transitions are indicated by the sharp peaks. Notably, bound-free
transition vanishes at $a'=a$.} \label{fig4}
\end{figure}

Figure~\ref{fig4} shows a continuous change near a Feshbach
resonance from a bound-free lineshape for $a'<0$ to a
combination of bound-bound and bound-free transitions for $a'>0$.
When the scattering length approaches negative
infinity, the linewidth of the bound-free lineshape approaches
zero as $\sim E_{b'}^2\sim {a'}^{-4}$ and ``evolves'' into the
bound-bound delta function. Remarkably, for $a'>0$, the bound-free
transition is much weaker. At $a'=a$, the bound-free transition is
fully suppressed, $\Gamma_f=0$.

\begin{figure}
\includegraphics[width=2.75in]{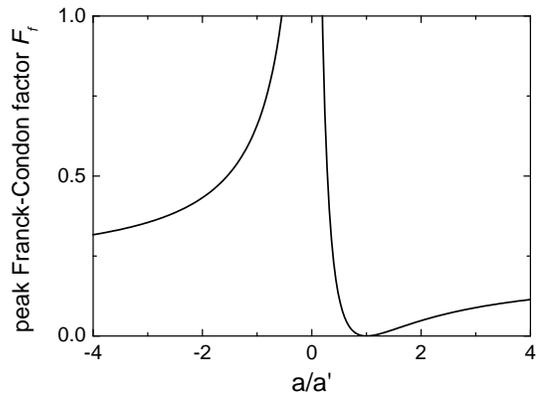}
\caption{Fano profile near a Feshbach resonance in the outgoing
channel $A'$. The peak Franck-Condon factors with $E_b=1$ are
calculated for different $a/a'$ based on Eq.~(\ref{ffkfinal}).
Near the Feshbach resonance, $a/a'$ approaches zero.} \label{fig5}
\end{figure}

The vanishing bound-free transition for $a'\approx a$ can be
understood by the wavefunction overlap given in Eq.~(\ref{ffk}).
As a positive scattering length $a'$ indicate a zero in
wavefunction near $r\approx a$, the sign change of the scattering
wavefunction results in a cancellation in the overlap.
Alternatively, we may consider the transition amplitude to the
continuum $A'$ is exactly cancelled by the Feshbach coupling
between molecular state $|m'\rangle$ and the continuum $A'$. This
interference effect results in the Fano-like profile of the peak
transition rate near a Feshbach resonance \cite{fano} (see
Fig.~\ref{fig5}).

On the other hand, the bound-bound Franck-Condon factor approaches
one when $a'\approx a$, since both molecular
wavefunction are identical for $r>r_0$. The peak bound-free transition rates
therefore shows a strong asymmetry with respect to the sign of the
scattering length, shown in Fig.~\ref{fig5}.

Based on the above features, the rf spectroscopy does provides a
new strategy to extract various important scattering parameters.
In particular, while cold collision measurements are generally
insensitive to the sign of the scattering length, our rf
excitation spectroscopy is drastically different for the two cases
and provides a new tool to probe the scattering properties.

\section{Comparison with experiment}

To compare with the experimental result, we take $^6$Li as a model
system and calculate the rf spectra. The experimental spectra are
obtained from the Li group in Innsbruck \cite{boundbound}.

Adopting the convention used in Ref.\cite{gap}, we define the rf
offset energy as $E=E_{RF}-E_0=K+E_b$. Using Eq.~(\ref{ffkfinal}),
we can rewrite the Franck-Condon factors as

\begin{eqnarray}
F_f(E)&=&\frac{2}{\pi}(1-\frac{a'}{a})^2\frac{E_b'E_b^{1/2}(E-E_b)^{1/2}}{E^2(E+E_b'-E_b)} \label{expfcf1}\\
F_b(E)&=&\frac{4aa'}{a+a'}\delta(E-E_b+E_b') \mbox{ for } a'>0
\label{expfcf2}
\end{eqnarray}

\begin{figure}
\includegraphics[width=2.75in]{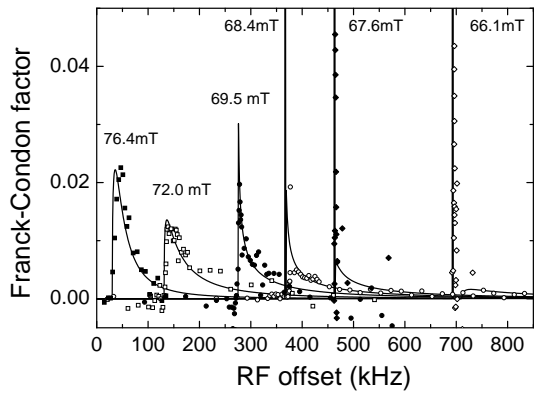}
\caption{RF spectra of $^6$Li$_2$ molecule at different magnetic
field values. Theoretical curves are based on Eq.~(\ref{expfcf1})
and Eq.~(\ref{expfcf2}) with the binding energies and scattering
lengths obtained from the multichannel calculation. To shown both
the bound-free and bound-bound spectra, an integration bandwidth
of 1kHz is assumed. RF-induced loss on the molecular population at
76.4mT (solid squares), 72.0mT (open squares), 69.5mT (solid
circles), 68.4mT (open circles), 67.6mT (solid diamonds) and
66.1mT (open diamonds) are re-scaled. Experiment data are from the
Grimm's group in Innsbruck.} \label{fig6}
\end{figure}

We compare the theoretical curves with the experiment in
Fig.~\ref{fig6}. An integration bandwidth of 1kHz is assumed. When
the magnetic field approaches 69.1mT from high values, the
molecular binding energy $E_b$ in the $(1,2)$ channel increases
(see Fig.~\ref{fig1}). This dependence is shown in Fig.~\ref{fig6}
as the whole excitation line moves toward higher frequency. At the
same time, the lineshape becomes sharper due to the Feshbach
resonance in $(1,3)$ channel at 69.1mT. Below 69.1mT, we expect
both bound-free and bound-bound transitions are allowed. At
68.4mT, these two components coexist, but cannot be clearly
distinguished due to the small binding energy of $E_b'=h\times
0.5$kHz, lower than the experimental resolution of $\sim 1$kHz. At
67.6mT and 66.1mT, the bound-free transitions are strongly
suppressed. The bound-bound transition then shows up as the
dominant component in the spectra. The excellent agreement between
the experiment and the calculation over a large range of magnetic
fields is remarkable. This result strongly suggests that the rf
molecular spectroscopy can be used for a precision determination
of the atomic interaction parameters \cite{boundbound}.

\section{Conclusion}
We model and evaluate the radio-frequency excitation rates on
weakly-bound ultracold molecules in the threshold regime. We
derive a simple and analytic form of the bound-free and
bound-bound spectral profiles which provide an excellent fit to
both the numerical calculation and the recent rf measurements on
Li$_2$ molecules.

An interesting case is studied when a Feshbach resonance occurs in
the outgoing channel. We show that the bound-free spectra in the
absence of a bound state smoothly evolves into a combination of
bound-bound and bound-free spectra when the bound state formed
near a Feshbach resonance. The bound-free transition rate strongly
depends on the sign of the scattering length and shows a Fano-like
structure near the resonance.

We like to point out that the rf spectroscopy based on
weakly-bound molecules can be an excellent tool to determine the
long-range interaction properties with high precision. From the
excitation spectra, the molecular binding energies, the atomic
scattering lengths and their signs, and the scattering phase
shifts can be determined. Furthermore, in contrast to conventional
cold collision measurements, rf transitions on molecules are
insensitive to the sample density and temperature, can probe
different scattering channels by tuning the rf frequency to
different states, and can be implemented instantly without
thermalization. Finally, rf transitions between molecular states
may also provide a new avenue to transfer the molecular population
to low-lying molecular states.

\section*{Acknowledgements}
We thank R. Grimm's Li group in Innsbruck for providing the
experimental data and stimulating discussions. P.S. Julienne would
like to thank the Office of Naval Research for partial support.
C.C. is a Lise-Meitner research fellow of the Austrian Science
Fund.


\end{document}